\documentclass{jpp}
\usepackage{graphicx}
\usepackage{lineno}
\usepackage{hyperref}

\usepackage[utf8]{inputenc}
\usepackage[T1]{fontenc}
\usepackage{amsmath}

\renewcommand{\vec}[1]{\boldsymbol{#1}}

\shorttitle{Velocity shears in the solar wind}
\shortauthor{Simon Opie et al.}

\title{Temperature anisotropy instabilities driven by intermittent velocity shears in the solar wind}

\author{Simon Opie\aff{1}
 \corresp{\email{simon.opie.18@ucl.ac.uk}},
 Daniel Verscharen\aff{1}
 Christopher H. K. Chen\aff{2}
 Christopher J. Owen\aff{1}
 Philip A. Isenberg\aff{3}
 Luca Sorriso-Valvo\aff{4,5} 
 Luca Franci\aff{6}
 Lorenzo Matteini\aff{7}}

\affiliation{\aff{1}Mullard Space Science Laboratory, University College London, Dorking, RH5 6NT, UK 
\aff{2}Department of Physics and Astronomy, Queen Mary University of London, London, E1 4NS, UK \aff{3} Space Science Center, University of New Hampshire, Durham, NH 03824, USA \aff{4} CNR, Istituto per la Scienza e la Tecnologia dei Plasmi, via Amendola 122/D, 70126 Bari, Italy \aff{5} Space and Plasma Physics, School of Electrical Engineering and Computer Science, KTH Royal Institute of Technology, Teknikringen 31, 11428 Stockholm, Sweden  \aff{6} Department of Mathematics, Physics and Electrical Engineering, Northumbria University, Newcastle upon Tyne, NE1 8ST, UK \aff{7} The Blackett Laboratory, Department of Physics, Imperial College London, London, SW7 2AZ, UK}

\begin{document}

\maketitle

\begin{abstract}
Where and under what conditions the transfer of energy between electromagnetic fields and particles takes place in the solar wind remains an open question. We investigate the conditions that promote the growth of kinetic instabilities predicted by linear theory, to infer how turbulence and temperature-anisotropy-driven instabilities are interrelated. Using a large dataset from Solar Orbiter, we introduce the radial rate of strain, a novel measure computed from single-spacecraft data, that we interpret as a proxy for the double-adiabatic strain rate. The solar wind exhibits high absolute values of the radial rate of strain at locations with large temperature anisotropy. We measure the kurtosis and skewness of the radial rate of strain from the statistical moments to show that it is non-Gaussian for unstable intervals and increasingly intermittent at smaller scales with a power-law scaling. We conclude that the velocity field fluctuations in the solar wind contribute to the presence of temperature anisotropy sufficient to create potentially unstable conditions. 

\end{abstract}

\section{Introduction}

The solar wind is a weakly collisional, magnetised plasma characterised by kinetic processes that influence its dynamic evolution in ways that are not fully understood. The  expansion of the solar wind into interplanetary space in the presence of a decreasing background magnetic field  \citep{matteini_evolution_2007,matteini_ion_2012} implies that the particle distributions should be highly anisotropic by the time the plasma reaches a distance of $\sim1\,\mathrm{au}$ from the Sun \citep{verscharen_collisionless_2016}. The double-adiabatic \citep[CGL;][]{chew_boltzmann_1956} expansion of the solar wind predicts a decline in $T_\perp/T_\parallel$ with distance from the Sun, where $T_\perp$ ($T_\parallel$) is the proton temperature perpendicular (parallel) to the magnetic field direction \citep{matteini_evolution_2007,cranmer_empirical_2009}. However, observations at $1\,\mathrm{au}$ show that the solar wind is on average almost isotropic with respect to the background magnetic field, albeit with significant variability in temperature anisotropy about the isotropic equilibrium \citep{marsch_solar_1982,kasper_windswe_2002,bale_magnetic_2009, maruca_what_2011,isenberg_self-consistent_2013,coburn_measurement_2022}.  

The presence of temperature anisotropy in the solar wind is also linked to turbulence. Solar-wind turbulence facilitates a non-linear transfer of energy from larger to smaller scales via a Kolmogorov-like inertial range, leading to the dissipation of energy at kinetic scales  \citep{kolmogorov_dissipation_1941, alexandrova_solar_2013,bruno_solar_2013,kiyani_dissipation_2015, chen_recent_2016,marino_scaling_2023}. 
This non-linear energy transfer occurs via a cascade which is inherently anisotropic in the distribution of its spectral power with respect to the wavevector of the fluctuations, with $k_\perp~\gg~k_\parallel$, where $k_\perp$ ($k_\parallel$) is the component of the wavevector perpendicular (parallel) to the magnetic field direction \citep{cho_anisotropy_2000,schekochihin_astrophysical_2009,wicks_anisotropy_2011, horbury_anisotropy_2012,oughton_anisotropy_2015,chen_multi-species_2016,schekochihin_mhd_2022}. Some solar wind models predict that the turbulent cascade is responsible for the temperature-anisotropic heating of the plasma \citep{parashar_kinetic_2009,chandran_perpendicular_2010, howes_dynamical_2015}. 
Considering the expectations based on CGL expansion and temperature-anisotropic heating models, the observation of approximately isotropic plasma conditions on average suggests that additional processes act to restore isotropy through the transfer of energy \citep{marsch_solar_1982}. The stability of the solar wind depends on the simultaneous contributions of all species in the plasma to its free energy \citep{chen_multi-species_2016} but here we consider only the protons.

A class of kinetic instabilities is triggered when the proton temperature anisotropy exceeds certain thresholds for the production of plasma waves and non-propagating modes. These instabilities transfer energy from the particles to electromagnetic fields and this transfer restores the proton distribution towards a stable state closer to isotropy \citep{gary_proton_1976,gary_theory_1993, hellinger_solar_2006}.
At large scales, compressive fluctuations can drive temperature anisotropy leading to wave-driven instabilities that eventually reduce anisotropy through pitch-angle scattering of protons \citep{verscharen_collisionless_2016}, while at small scales kinetic instabilities predicted by linear theory, redistribute energy through wave--particle interaction \citep{gary_theory_1993,kasper_windswe_2002,hellinger_heating_2011,hellinger_proton_2013,howes_dynamical_2015}. 

Proton-kinetic processes, such as temperature-anisotropy-driven instabilities, predominantly occur on scales near the small-scale end of the inertial range \citep{gary_short-wavelength_2015}. Contrary to the framework of traditional linear theory, kinetic instabilities in the solar wind operate in inhomogeneous  and non-constant conditions due to the ubiquitous solar-wind turbulence \citep{coleman_turbulence_1968,frisch_simple_1978,tu_mhd_1995,chen_recent_2016,verscharen_multi-scale_2019, opie_conditions_2022, opie_effect_2023}. In developing a more robust understanding of where and under what conditions energy transfer takes place, it is therefore important to fully capture the interplay between kinetic and turbulent features at the appropriate scales in the solar wind \citep{osman_kinetic_2012,chen_recent_2016,sorriso-valvo_statistical_2018,opie_conditions_2022,opie_effect_2023,arzamasskiy_kinetic_2023}.

The solar-wind turbulence is intermittent and consistent with the model of multifractality  \citep{frisch_turbulence_1995}, meaning that the fluctuations at different scales are not equally space filling and instead contain coherent structures such as current sheets and velocity shear layers \citep{sorriso-valvo_intermittency_1999,greco_intermittent_2008,osman_proton_2013,servidio_proton_2014,matthaeus_intermittency_2015,qudsi_observations_2020}. These structures extend across scales in the inertial range and exhibit a statistical scaling relationship that indicates that they are self-affine \citep{carbone_experimental_1995,carbone_evidences_1996,sorriso-valvo_intermittency_1999,kiyani_extracting_2006,hnat_fractal_2007}. 

We hypothesise that the turbulent and intermittent velocity field in the solar wind is dynamically important for driving the temperature anisotropy of the plasma protons. Using a large observational dataset, we localise conditions in the solar wind at and beyond the thresholds for the proton-driven oblique firehose instability when $T_\perp/T_\parallel<1$ \citep{hellinger_oblique_2008,markovskii_effect_2022} and mirror-mode instability when $T_\perp/T_\parallel>1$ \citep{kunz_firehose_2014,hellinger_mirror_2017}, which place effective boundaries for temperature anisotropy in the solar wind \citep{hellinger_solar_2006,bale_magnetic_2009,gary_short-wavelength_2015}. Working directly from the dynamical equations, we develop and analyse a quantitative measure for the impact of velocity shears on the temperature anisotropy, the radial rate of strain. We measure the third and fourth statistical moments of the radial rate of strain, the velocity field, and the magnetic field, from which we infer where, in terms of turbulent structures in the solar wind, these instabilities are located. 

We set out details of our data analysis in Section~\ref{DAna}. In Section~\ref{RadS}, we develop and evaluate our novel measure of the radial rate of strain, a one-dimensional proxy for the three-dimensional double-adiabatic strain rate. In Section~\ref{intsec2}, we calculate the skewness and kurtosis of the radial rate of strain, the magnetic field and the velocity field. We discuss the significance of our results in Section~\ref{disc} and conclude with recommendations for further work in Section~\ref{conc}.

\section{Data Analysis}
\label{DAna}

\subsection{Dataset}

\begin{table}
  \begin{center}
\def~{\hphantom{0}}
  \begin{tabular}{lcc}
      Interval  &   Heliocentric Distance ($R_S$)  &  Number of Datapoints\\[3pt]
        2020 October 07-18 & 205 & 185\,923 \\
        2021 April 22-28 & 190 & 131\,481\\
        2021 May 05-11 & 180 & 131\,849\\
        2021 June 10-13 & 200 & 79\,641\\
        2021 July 06-11 & 190 & 117\,427\\
        2021 July 20-24 & 180 & 88\,429\\
        2021 October 09-12 & 150 & 81\,362\\
        2021 October 19-26 & 160 & 159\,404\\
        2021 November 02-07 & 175 & 132\,750\\
        2021 November 09-14 & 180 & 86\,201\\
        2021 November 16-19 & 190 & 85\,833\\
        2021 December 28 - 2022 January 02 & 215 & 128\,048\\
        2022 January 04-08 & 215 & 58\,773\\
  \end{tabular}
  \caption{Data selection from the Solar Orbiter Archive with approximate heliocentric distance (in solar radii, $R_S$) for each data interval}
  \label{tab:dset}
  \end{center}
\end{table}

Our dataset, which is a significant extension of that used by \citet{opie_conditions_2022},  comprises $\approx1.5\mathrm{M}$ datapoints, as detailed in Table~\ref{tab:dset}. Our data are taken from the Solar Orbiter public archive\footnote{\url{http://soar.esac.esa.int/soar/}}. We use data from two of the in-situ instruments on board the spacecraft which make measurements of the solar wind; namely the magnetic-field vector from the Magnetometer (MAG) at $8\,\mathrm{Hz}$ cadence \citep{horbury_solar_2020} and the proton moments from the Proton Alpha Sensor (PAS). For the periods discussed here, PAS takes a $1\,\mathrm{s}$ sample every $4\,\mathrm{s}$. PAS is part of the Solar Wind Analyser (SWA) instrument suite  \citep{owen_solar_2020}. For this study, we work in the $(R,T,N)$ coordinate system, where the axis $R$ points radially outwards from the Sun, $T$ is given by the cross-product between the Sun's rotation vector and $R$, and $N$ completes the right-handed triad.  

The relative error for the PAS data is $\approx0.27\%$ for the velocity measurements \citep{louarn_multiscale_2021}. For our dataset, this represents an average absolute error of $\approx1.2\,\mathrm{kms^{-1}}$.  

Since we employ two-point field increments in this analysis, we use continuous data intervals of 4--7 days duration, subject to data availability. We exclude PAS datapoints for which the solar wind bulk velocity $<325\,\mathrm{km}\,\mathrm{s}^{-1}$ or when measurements are outside the recommended PAS quality factor $\leq0.2$. No attempt is made to eliminate any structures, such as shocks or interplanetary coronal mass ejections (ICMEs), from the dataset. Reference to the Helio4Cast catalogue\footnote{\url{https://helioforecast.space/static/sync/icmecat/HELIO4CAST\_ICMECAT\_v22.csv}} for ICMEs observed by Solar Orbiter gives three ICMEs in total within our combined dataset. These occurred on 2021 May 06, 2021 May 10, and 2021 November 03 for an aggregated duration of 79.16 hours of our total observational timeframe of 80 days \citep{mostl_modeling_2017,mostl_prediction_2020}.

Solar Orbiter has provided a continuous high-resolution dataset of the pristine solar wind for both magnetic and proton velocity field analysis \citep{louarn_multiscale_2021, zouganelis2022solar}. This enables our two-point field increment method to be  statistically robust on timescales that equivalent to the scales needed to observe unstable intervals \citep{opie_conditions_2022}.

\subsection{Data processing}

We rotate the proton pressure tensor to align with the local associated magnetic field and create a timeseries for $\beta_{\parallel}\equiv 8\pi n_{\mathrm p}k_{\mathrm B}T_{\parallel}/B^2$, where $n_{\mathrm p}$ is the proton number density, $k_{\mathrm B}$ is the Boltzmann constant, and $B$ is the magnetic field strength averaged over the associated 1\,s PAS measurement interval.  From the proton pressure tensor, we then calculate the ratio $T_{\perp}/T_{\parallel}$ for each PAS measurement.

In order to identify intervals in our dataset for which linear theory predicts the growth of kinetic instabilities, we require thresholds for the growth of the specific anisotropy-driven kinetic instabilities of interest. We base our identification of unstable intervals on the parametric  approximation for the instability thresholds in the form
    \begin{equation} \label{ethresh}
        \frac{T_{\perp}}{T_{\parallel}} = 1 + \frac{a}{(\beta_{\parallel} - c)^b},
    \end{equation}
where $a$, $b$, and $c$ are constants with values given for each instability for a range of instability maximum growth rates by \citet{verscharen_collisionless_2016}. We use the constants given for a maximum growth rate of  $\gamma_{\mathrm m}  = 10^{-2} \Omega_{\mathrm p}$, where $\Omega_{\mathrm p}$ is the proton gyrofrequency. We evaluate these instability thresholds for the oblique firehose (OF) and for the mirror-mode (M) instabilities. For reference, we also include the instability threshold for the  Alfv\'en/ion-cyclotron (A/IC) instability in part of our analysis.

We analyse data distributed in anisotropy - plasma beta ($T_{\perp}/T_{\parallel}$--$\beta_{\parallel}$) parameter space that is bounded by these thresholds. We define the unstable intervals as comprising those data points that lie above the thresholds, while we characterise all data below the thresholds as stable. Consequently, the plasma is considered unstable against M and A/IC instabilities if $T_{\perp}/T_{\parallel}$ is greater than the value given by the right-hand side of Eq.~(\ref{ethresh}), while it is considered unstable against the OF instability if $T_{\perp}/T_{\parallel}$ is less than the value given by the right-hand side of Eq.~(\ref{ethresh}), with in each case the constants chosen for the respective instability.

\section{Radial strain}
\label{RadS}

In this section, we use a single-species (proton) fluid model of space plasma that derives anisotropy directly from the effect of fluid strain on the pressure tensor. We define the radial rate of strain $\Gamma_R$ as a proxy for the strain rate in fully three-dimensional turbulence for this analysis, which we believe to be a novel technique. We then use this measure to test our hypothesis that intermittent velocity shear is dynamically important for the generation of the observed temperature anisotropy in the solar wind.

\subsection{Method: Determination of the radial strain}

We examine the evolution of pressure anisotropy based on the first three moments of the proton distribution function under CGL analysis \citep{chew_boltzmann_1956, kulsrud_mhd_1984}. We take the form of the equation for pressure anisotropy given by \citet{squire_pressure_2023} which assumes that the pressure tensor is invariant to rotation about the magnetic field direction:
\begin{multline}
    \frac{\mathrm d}{\mathrm dt}(p_{\perp}-p_{\parallel})= (p_{\perp} + 2p_{\parallel})\boldsymbol{\hat{b}}\boldsymbol{\hat{b}}\boldsymbol{:}\nabla\boldsymbol{v}- (2p_{\perp} - p_{\parallel})\nabla\bcdot\boldsymbol{v}\\
    -\nabla\bcdot[(q_{\perp}-q_{\parallel})\boldsymbol{\hat{b}}]-3q_{\perp}\nabla\bcdot\boldsymbol{\hat{b}} - \nu_{\mathrm c}(p_{\perp}-p_{\parallel}),
    \label{Panis}
\end{multline}
where $p_{\perp}$ and $p_{\parallel}$ are the proton pressure tensor components perpendicular and parallel to the magnetic field, $\boldsymbol{\hat{b}}=\boldsymbol{B}/B$, $\vec {v}$ is the proton bulk velocity, $q_\perp$ and $q_\parallel$ are the proton heat fluxes perpendicular and parallel to the magnetic field direction, and $\nu_{\mathrm c}$ is the proton collisional relaxation frequency for temperature anisotropy. The first two terms on the right-hand side of Equation~(\ref{Panis}) describe how plasma bulk flows directly affect pressure anisotropy through shear and compression. We assume that $\nu_{\mathrm c}=0$ and  $q_\perp$=$q_\parallel$=$0$  following the arguments of \citet{del_sarto_shear-induced_2018}. 

These assumptions allow us to formulate a new proxy measure for the driving of pressure anisotropy by plasma motion based on plasma measurements sampled along the radial direction by a single spacecraft. We use Taylor's hypothesis \citep{taylor_spectrum_1938} to transpose spatial partial derivatives to temporal partial derivatives and employ increments to replace $\nabla\bcdot\boldsymbol{v}$ and $\nabla\boldsymbol{v}$. After these transformations, we define the radial rate of strain as 
\begin{equation}
\Gamma_R = \frac{1}{\tau}\left[\left[(p_\perp + 2p_\parallel)\left({\hat{b}_R}{\hat{b}_R}\Delta{v_R}+{\hat{b}_T}{\hat{b}_R}\Delta{v_T}+{\hat{b}_N}{\hat{b}_R}\Delta{v_N}\right)\right] -  (2p_\perp - p_\parallel)\Delta{v_R}\right],
\label{RRS}
\end{equation}
where $\tau$ is the time increment between measurement points (i.e., the measurement cadence) and $\Delta\phi$ denotes the scale-dependent increment $\Delta\phi=\phi(t)-\phi(t+\tau)$ of any time-dependent observable $\phi$ in the spacecraft reference frame.

Following the interpretation of the first and second terms in Equation~\eqref{Panis}, we decompose Equation~(\ref{RRS}) into proxies for the incompressible ($\Gamma_{R\mathrm I}$)  and compressible ($\Gamma_{R\mathrm C}$) contributions to $\Gamma_R$ as
\begin{equation}
    \label{RRSI}
    \Gamma_{R\mathrm I} = \frac{1}{\tau}\left[(p_\perp + 2p_\parallel)\left({\hat{b}_R}{\hat{b}_R}\Delta{v_R}+{\hat{b}_T}{\hat{b}_R}\Delta{v_T}+{\hat{b}_N}{\hat{b}_R}\Delta{v_N}\right)\right]
\end{equation}
and
\begin{equation}
    \label{RRSC}
    \Gamma_{R\mathrm C} = \frac{1}{\tau}\left[(p_\parallel - 2p_\perp)\Delta{v_R}\right].
\end{equation}
We set $\tau=8\,\mathrm{s}$, which results in an average relative value of $\Delta{v}_j/|\boldsymbol{v}|$ for our intervals of $0.48\%$. We check that all data with small $\Delta{v}_j$ values make a negligible contribution to $\Gamma_R$ and so the relevant datapoints are well above the PAS relative error. Any systematic errors due to bias offsets in any quantity $\phi$ are eliminated through the process of calculating the difference $\Delta\phi$. We exclude any data points for which an increment of $\tau=8\,\mathrm{s}$ cannot be calculated due to data gaps.

\begin{figure}
  \centering
  \includegraphics{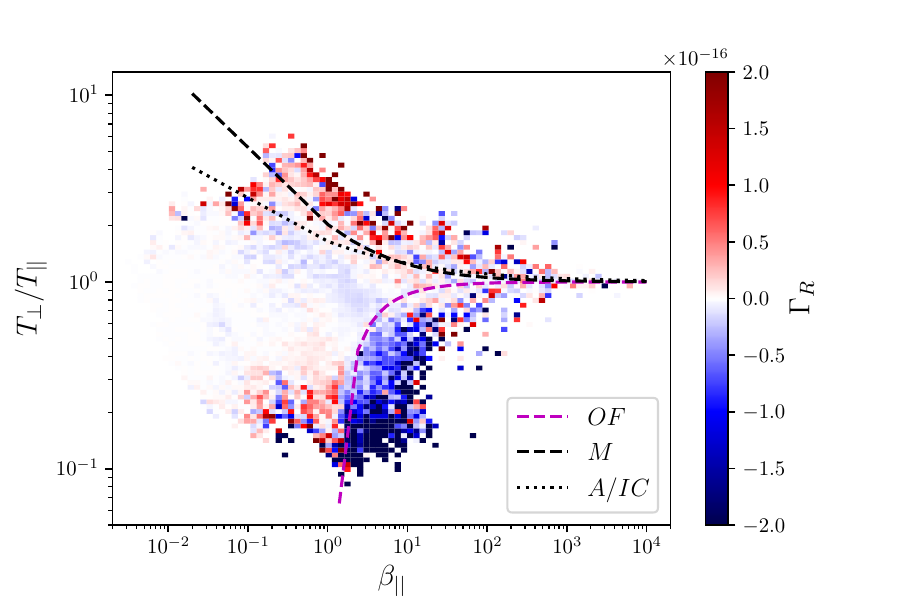}
  \caption{Distribution of $\Gamma_R$  as bin averages in $T_{\perp}/T_{\parallel}$--$\beta_{\parallel}$ parameter space. We overplot the  instability thresholds  for the Oblique Firehose (OF), Alfv\'en/Ion cyclotron (A/IC), and Mirror-mode (M) instabilities.}
\label{fig:rad1}
\end{figure}

\begin{figure}
  \centering
  \includegraphics{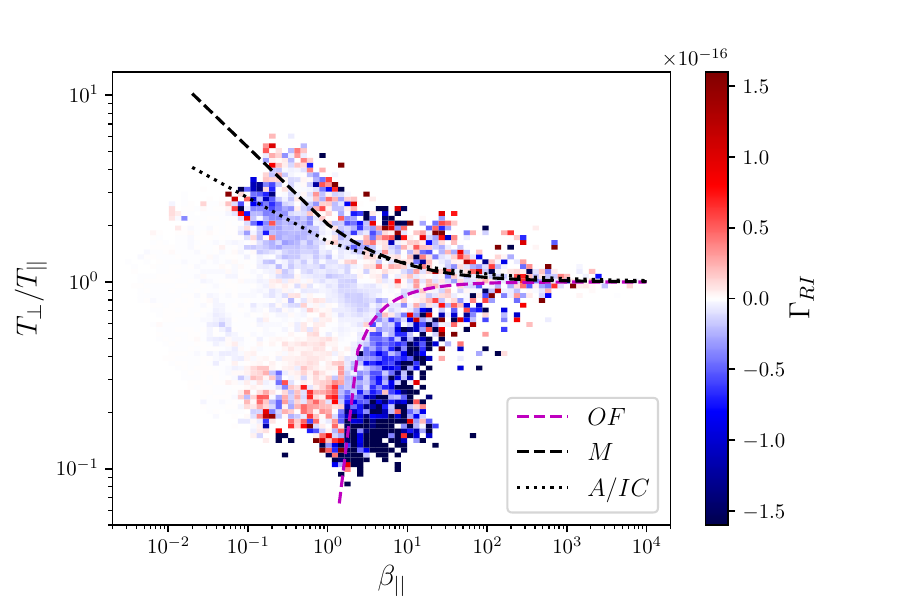}
  \caption{Distribution of $\Gamma_{R\mathrm I}$ as bin averages in $T_{\perp}/T_{\parallel}$--$\beta_{\parallel}$ parameter space. We overplot the instability thresholds  for the Oblique Firehose (OF), Alfv\'en/Ion cyclotron (A/IC), and Mirror-mode (M) instabilities.}
\label{fig:rad2}
\end{figure}

\begin{figure}
  \centering
  \includegraphics{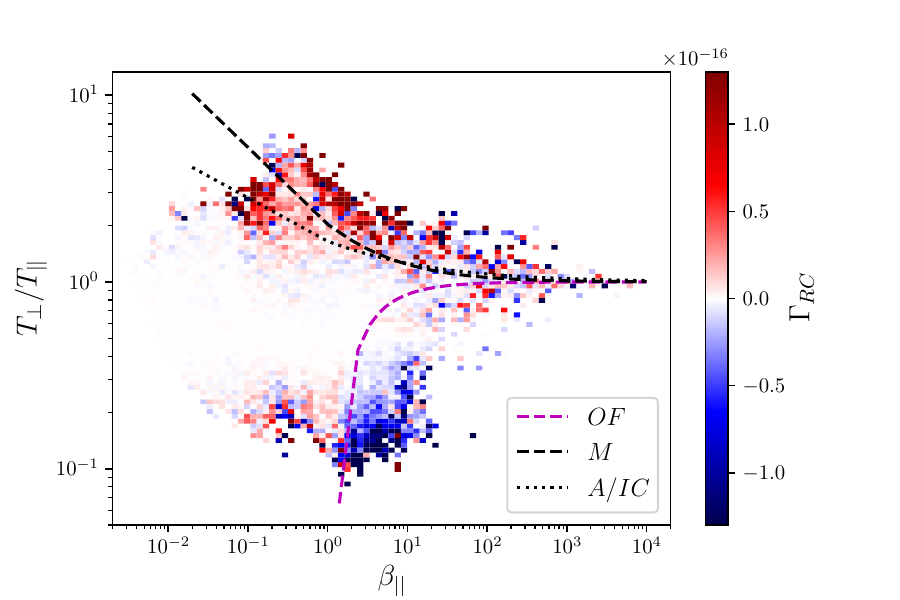}
  \caption{Distribution of $\Gamma_{R\mathrm C}$ as bin averages in $T_{\perp}/T_{\parallel}$--$\beta_{\parallel}$ parameter space. We overplot the instability thresholds  for the Oblique Firehose (OF), Alfv\'en/Ion cyclotron (A/IC), and Mirror-mode (M) instabilities.}
\label{fig:rad3}
\end{figure}

\subsection{Results: The radial strain of stable and unstable plasma intervals}

We calculate $\Gamma_R$ from Equation~\eqref{RRS} for each datapoint in our dataset after the elimination of data gaps and plot the bin-averaged values in a two-dimensional histogram in $T_{\perp}/T_{\parallel}$--$\beta_{\parallel}$  parameter space in Figure~\ref{fig:rad1}. We also calculate $\Gamma_{R\mathrm I}$ from Equation~\eqref{RRSI} and $\Gamma_{R\mathrm C}$ from Equation~\eqref{RRSC} and plot these separately in two-dimensional histograms in $T_{\perp}/T_{\parallel}$--$\beta_{\parallel}$ parameter space in Figures~\ref{fig:rad2} and \ref{fig:rad3}, respectively. 

According to Figure~\ref{fig:rad1}, the unstable intervals show on average high absolute values, relative to the stable data, of $\Gamma_R$, with $\Gamma_R>0$ in the parameter space associated with the M and A/IC instabilities and $\Gamma_R<0$ in the parameter space associated with the OF instability. This relation is equally maintained for both $\Gamma_{R\mathrm I}$ and $\Gamma_{R\mathrm C}$ as shown in Figures~\ref{fig:rad2} and \ref{fig:rad3}, although the signal varies with $\beta_{\parallel}$ and $T_{\perp}/T_{\parallel}$. The contribution from $\Gamma_{R\mathrm C}$ is important at lower $\beta_{\parallel}$ and higher $T_{\perp}/T_{\parallel}$ for the mirror-mode unstable region of parameter space and at lower values of $T_{\perp}/T_{\parallel}$ for the oblique firehose unstable region. For the stable data, the average value of $\Gamma_R$ is $\sim 0$ except for the boundary regions of stable parameter space where either $T_{\perp}/T_{\parallel}<0.3$ or $T_{\perp}/T_{\parallel}>2$. This result is consistent for both $\Gamma_{R\mathrm I}$ and $\Gamma_{R\mathrm C}$, although we see some variance in the sign of $\Gamma_{R\mathrm I}$ at the boundary to the mirror-mode unstable region. 

We verify from the probability density functions (not shown here) of $\Gamma_R$ for each of the stable ($\Gamma_R^S$), oblique firehose unstable ($\Gamma_R^{OF}$), and mirror-mode unstable ($\Gamma_R^M$) regions that the magnitude of the averaged $\Gamma_R$ is consistently larger in the unstable measurement intervals and that the contrast with the stable region is not simply an averaging effect. We find the mean values  as $\langle\Gamma_R^S\rangle=-3.76\times10^{-18}\,\mathrm{J\,cm^3\,s^{-1}}$, $\langle\Gamma_R^{OF}\rangle=-4.95\times10^{-17}\,\mathrm{J\,cm^3\,s^{-1}}$ and $\langle\Gamma_R^M\rangle=3.51\times10^{-17}\,\mathrm{J\,cm^3\,s^{-1}}$, where $\langle\cdot\rangle$ represents the average over all measurement intervals in the respective category.

\section{Intermittency}
\label{intsec2}

In this section, we analyse  statistical moments, specifically the third (skewness) and fourth (kurtosis) moments, to measure the intermittency of $\Gamma_R$. We compare this result with the skewness and kurtosis of the background magnetic and velocity fields.

Intermittency of magnetic and electric fields and plasma parameters (velocity, density, temperature) in the inertial range has been extensively studied \citep{tu_mhd_1995,marsch_intermittency_1997,sorriso-valvo_intermittency_1999,bruno_identifying_2001}. We examine the statistical measures separately for stable and unstable intervals to understand the dynamics of the driving of anisotropy by the turbulent fluctuations. In particular, we investigate how the intermittency (in the sense of `burstiness') of the pressure strain varies across the relevant scales.

\subsection{Method: Measuring the Skew and Kurtosis across scales}

We use the statistical moments of the quantities $\mu_i$ to calculate their skewness $\lambda_i$ and their kurtosis $\kappa_i$, where $\mu_i$ is the moment about the mean. The moments are defined as 
\begin{equation}
    \lambda_i = \frac{\langle\mu_i^3\rangle}{\langle\mu_i^2\rangle^{3/2}}.
    \label{eqint7}
\end{equation}
and 
\begin{equation}
    \kappa_i = \frac{\langle\mu_i^4\rangle}{\langle\mu_i^2\rangle^2}
    \label{eqint5}
\end{equation}

Using Equations~\ref{eqint7} and \ref{eqint5}, we calculate $\lambda_i$ and $\kappa_i$ for $\Gamma_R$, $\boldsymbol{B}$ and $\boldsymbol{V}$. 
We calculate $\lambda_i$ and $\kappa_i$  for the whole dataset and for each of the stable, oblique firehose unstable, and mirror-mode unstable datasets.

We also calculate $\lambda_{\Gamma_R}$ and $\kappa_{\Gamma_R}$  across a range of temporal scales. We do this by creating scale-dependent datasets for $\Gamma_R$ using Equation~(\ref{RRS}) with $\tau\in\{8\,\mathrm{s}, 16\,\mathrm{s}, 32\,\mathrm{s}, 64\,\mathrm{s}, 128\,\mathrm{s}, 256\,\mathrm{s}, 512\,\mathrm{s}, 1024\,\mathrm{s}, 2048\,\mathrm{s}\}$. This results in scale-dependent values of $\lambda_{\Gamma_R}$ and $\kappa_{\Gamma_R}$  from the small-scale end of the inertial range up to the correlation-length scale.

\subsection{Results: The intermittency of \texorpdfstring{$\mathrm{\Gamma}_R$}{}}

\begin{table}
  \begin{center}
\def~{\hphantom{0}}
  \begin{tabular}{lllllllll}
        & $\kappa_i$ & $\kappa^S_i$ & $\kappa^{OF}_i$ & $\kappa^M_i$ & $\lambda_i$ & $\lambda^S_i$ & $\lambda^{OF}_i$ & $\lambda^M_i$ \\[3pt]
        $\Gamma_R$ & 6.97e4 & 4.78e3 & 1.65e4 & 4.17e3 & -134 & -4.03 & -117 & +61.7 \\
        $\boldsymbol{V}$ & 4.78 & 4.84 & 2.69 & 2.43 & +1.05 & +1.07 & +0.31 & +0.602 \\
        $\boldsymbol{B}$ & 16.0 & 15.8 & 3.26 & 4.43 & +2.86 & +2.86 & +0.389 & +0.816 \\
        
  \end{tabular}
  \caption{ Skewness $\lambda_i$ and kurtosis $\kappa_i$ for $\Gamma_R$, $\boldsymbol{V}$, and $\boldsymbol{B}$ across all data, stable intervals, and unstable intervals.}
  \label{tab:interm}
  \end{center}
\end{table}

We show the values obtained for $\lambda_i$ and $\kappa_i$ in Table~\ref{tab:interm}.
The values of $\lambda_i$ indicate a symmetric distribution in nearly all cases except for $\lambda^{OF}_{\Gamma_R}$ which is strongly negative and $\lambda^{M}_{\Gamma_R}$ which is strongly positive, while the overall $\lambda_{\Gamma_R}$ is also strongly negative.

We see that $\kappa_{\Gamma_R} \gg \kappa_{\boldsymbol{V}}$ and $\kappa_{\Gamma_R} \gg \kappa_{\boldsymbol{B}}$; in both cases by $2-3$ orders of magnitude. This is expected since $\Gamma_R$ is built by multiplying intermittent quantities, and its normalised moments will therefore correspond to higher-order moments when compared with $\boldsymbol{V}$ and $\boldsymbol{B}$. For the conditioned subsets of $\Gamma_R$, $\kappa^{OF}_{\Gamma_R} > \kappa^{S}_{\Gamma_R} \approx \kappa^{M}_{\Gamma_R}$.

We calculate $\lambda_i$ and $\kappa_i$ for $\Gamma_{RI}$ and $\Gamma_{RC}$. $\lambda_{\Gamma_{RI}}\approx \mathrm{-104}$ is  negative  while $\lambda_{\Gamma_{RC}} \approx \mathrm{+46}$ is  positive. $\kappa_{\Gamma_{RI}}\approx \mathrm{5.1e4}$ and $\kappa_{\Gamma_{RC}} \approx \mathrm{3.05e4}$ are consistent with $\kappa_{\Gamma_{R}}$. 

\begin{figure}
  \centering
  \includegraphics{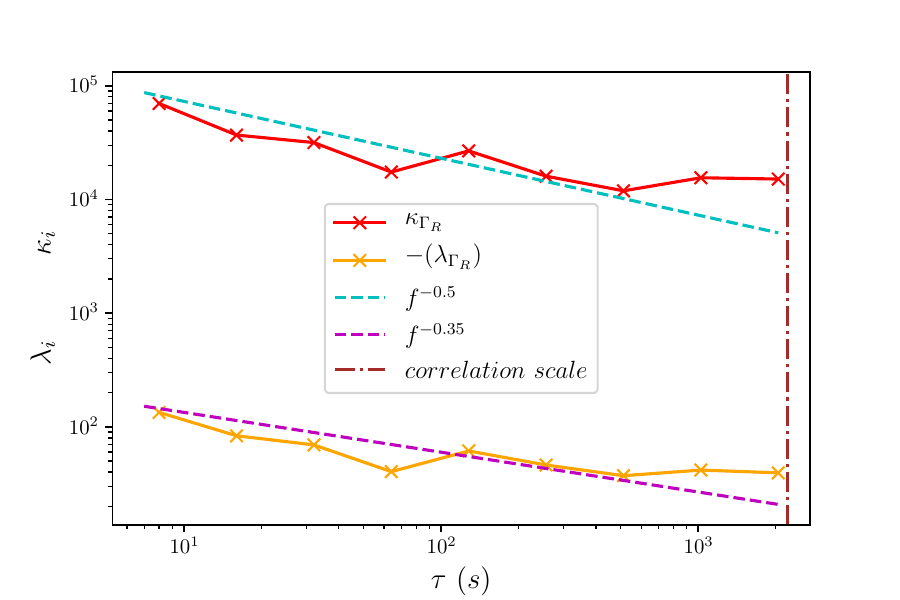}
  \caption{ $\lambda_{\Gamma_R}$ and $\kappa_{\Gamma_R}$  over a range of temporal scales $\tau$ from $8\,\mathrm s$ to $2048\,\mathrm s$. The cyan and magenta dashed lines represents a power-law scaling of $-0.5$ and $-0.35$ respectively. The brown vertical dash-dotted line represents the correlation scale.}
\label{fig:int3}
\end{figure}

Figure ~\ref{fig:int3} shows the scale-dependent values of $\lambda_{\Gamma_R}$ and $\kappa_{\Gamma_R}$  as a function of $\tau$. Both $\lambda_{\Gamma_R}$ and $\kappa_{\Gamma_R}$   show a power-law scaling which is overlaid as indicated. 
The values for $\lambda_{\Gamma_R}$  show that the skewness of $\Gamma_R$ increases with decreasing scale from a local minimum at $\tau\approx512\,\mathrm{s}$. The sign of all values of $\lambda_{\Gamma_R}$ is negative and has been reversed for visualisation on the logarithmic axis.

The values of $\kappa_{\Gamma_R}$ indicate that the burstiness of the field increases with decreasing scale, which is the signature of scale-dependent statistics typical of intermittency. Over all $\tau$ in this figure,  the $\Gamma_R$ field is evidently intermittent and hence `bursty' due to a heavy-tailed distribution of $\Gamma_R$. The value of $\kappa_{\Gamma_R}$ reaches a local minimum at $\tau\approx512\,\mathrm{s}$.

\section{Discussion}
\label{disc}

Under the assumptions underlying Equation~(\ref{Panis}) in Section~\ref{RadS}, shear and compression create pressure anisotropy \citep{squire_pressure_2023}, and we interpret our findings as being consistent with that prediction.  Relative to the stable data, the absolute values of the bin-averaged $\Gamma_R$ for the unstable data are significantly elevated, on average by a factor of ten. Our interpretation for this observation is that the presence of data in unstable intervals requires the pressure strain from turbulence-driven shears and compressions to compete with the instabilities that act to restore the plasma towards isotropy. In effect, data can only reside in unstable conditions if the timescale associated with the pressure-strain driving of temperature anisotropy is sufficiently small compared to the relaxation time of the relevant instabilities.

Previous work demonstrates that temperature anisotropy and vorticity are well correlated in hybrid simulations of plasma turbulence \citep{franci_two-dimensional_2016}. Hybrid-kinetic simulations also show that oblique firehose and mirror-mode instabilities can be driven by a changing magnetic field in a shear flow \citep{kunz_firehose_2014}. Our results  suggest that the turbulent shears in the velocity field create and modulate temperature anisotropy through a purely dynamical process such as that proposed by \citet{del_sarto_shear-induced_2018} and observed in near-Earth space \citep{servidio_proton_2014, sorriso-valvo_turbulence-driven_2019}. In this scenario, the non-compressive turbulent fluctuations result in spatially intermittent velocity shears that modulate the frozen-in magnetic field and drive $T_{\perp}/T_{\parallel}\ne 1$ with a corresponding change in magnetic field strength \citep{del_sarto_pressure_2016,del_sarto_shear-induced_2018,squire_pressure_2023}.

We also see a role for compressive fluctuations in this scenario as seen in Figure~\ref{fig:rad3} where the distribution of $\Gamma_{RC}$ is similar to the distribution of $\Gamma_R$. Figure~\ref{fig:rad3} also shows that the contribution from compressive fluctuations to $\Gamma_R$ in unstable intervals has the highest absolute average values where $T_{\perp}/T_{\parallel}$ approaches its maximum (M) or minimum (OF) bounds. In particular, we infer from comparison with Figure~\ref{fig:rad2} that compressive fluctuations are important in driving temperature anisotropy that creates conditions unstable to the mirror-mode instability. This is consistent with previous findings concerning the distribution of magnetic field fluctuations in unstable intervals \citep{opie_effect_2023}. We expect the contribution from $\Gamma_{RC}$ to be less significant overall since compressive fluctuations only account for a minor fraction of the solar wind turbulent energy \citep{tu_mhd_1995,bruno_solar_2013,verscharen_multi-scale_2019,marino_scaling_2023}.

The role of $\Gamma_R$ in driving mirror-mode unstable conditions is further supported by the skewness of $\Gamma_R$ and its conditioned subsets where, as set out in Section~\ref{intsec2},  both $\lambda_{\Gamma_RC}$ and $\lambda^M_{\Gamma_R}$ are positive while $\lambda_i$ is negative for the other subsets or close to zero for $\lambda^S_{\Gamma_R}$.

In the solar wind, the instabilities once triggered, act on timescales that rapidly reduce the temperature anisotropy \citep{bandyopadhyay_interplay_2022,opie_conditions_2022,opie_effect_2023}. The persistence of unstable intervals is, however, observable over extended spatial and temporal scales  \citep{opie_conditions_2022}. Although we cannot fully exclude the possibility that $\Gamma_R$ reflects the fluctuations created by the instabilities themselves rather than the anisotropy-driving background turbulence, this persistence time-scale argument suggests that the temperature anisotropies are indeed driven by the plasma motions. In addition to the consideration of the magnitude of $\Gamma_R$, an analysis of the normalised cross-helicity and the Alfv\'en ratio shows that there are structural differences in the turbulence in the unstable regions (see Appendix~\ref{appAlf}). This analysis suggests that there is an imbalance relative to the stable data whereby on average the energy in the velocity field is greater than that in the magnetic field for the unstable intervals. A further discussion of these structural differences would be worthwhile in a future study.

We note that our use of the radial rate of strain is distinct from the pressure strain interaction -- the so-called Pi-D formalism -- analysis that examines the contribution of turbulent fluctuations to the heat flux at ion scales \citep{yang_energy_2017,bandyopadhyay_statistics_2020,yang_quantifying_2023}. Nonetheless, our formulation goes some way to explaining the relationship between velocity shears and temperature anisotropy noted as an open question by \citet{yang_energy_2017}.

Similarly, our measure differs from the Partial Variation of Increments (PVI) which is used to locate magnetic (or velocity) field structures in solar-wind turbulence \citep{greco_intermittent_2008,greco_partial_2017}. The radial rate of strain $\Gamma_R$ is a dynamical measure that does not involve normalisation by a long-term average and is applied directly to fluctuations in the velocity field. We use $\Gamma_R$ to identify the creation of temperature anisotropy. We calculate the Pearson correlation coefficients between PVI and $\Gamma_R$ for our complete dataset and separately for each of the three conditioned subsets defined by stable, oblique firehose unstable, and mirror-mode unstable intervals. We find negligible to no correlation with coefficients that range from $\approx0.09$ for the stable data to $\approx0.003$ for the whole dataset.    

Observations and simulations suggest an important role for turbulent structures in the evolution of plasma conditions in the solar wind \citep{osman_kinetic_2012, servidio_proton_2014,franci_high-resolution_2015,greco_partial_2017,sorriso-valvo_multifractal_2017, hellinger_turbulence_2019, qudsi_observations_2020}. In Section~\ref{intsec2}, we use statistical analysis to link the role of $\Gamma_R$ in driving unstable conditions with the skewness and kurtosis of $\Gamma_R$, $\boldsymbol{B}$, and $\boldsymbol{V}$. We apply this analysis to the conditioned subsets denoted as stable, oblique firehose unstable, mirror-mode unstable. The large kurtosis of  $\Gamma_R$ and its subsets show  that $\Gamma_R$ is highly intermittent and exhibits burstiness to a greater degree than either $\boldsymbol{B}$ or $\boldsymbol{V}$. We see that $\kappa^{OF}_{\Gamma_R}$ is considerably greater than either $\kappa^{M}_{\Gamma_R}$ or $\kappa^{S}_{\Gamma_R}$ which is not the case for the subsets of $\boldsymbol{B}$ or $\boldsymbol{V}$.

Coupled with the values for the skewness, we find three distinct subsets of $\Gamma_R$ with positively and negatively skewed subsets being associated with the oblique firehose and mirror-mode instabilities respectively whilst for the stable data the skewness is nearly zero. The fact that the overall magnitudes of the skweness and kurtosis of $\Gamma_R$ are greater than those for the individual subsets implies that additional intermittency comes precisely from the alternation of these three distinct regions of stable data, firehose unstable data and mirror-mode unstable data.  

The power-law scalings of $\lambda_{\Gamma_R}$ and $\kappa_{\Gamma_R}$  shown in Figure~\ref{fig:int3} follow an exponent that is compatible with standard intermittency in the solar-wind magnetic field \citep{sorriso-valvo_intermittency_1999}. This correspondence is not unexpected since $\Gamma_R$ is built using various intermittent quantities. However, it implies that the greater magnitudes of $\lambda_{\Gamma_R}$ and $\kappa_{\Gamma_R}$  relative to those of $\boldsymbol{B}$ and $\boldsymbol{V}$ are not scale-dependent. We should therefore expect to see the regions of stable, oblique firehose unstable, and mirror-mode unstable data characterised by distinct subsets of $\Gamma_R$ across scales in the inertial range of solar-wind turbulence.

While it has long been known that the turbulence in the solar wind is intermittent and multifractal \citep{paladin_anomalous_1987,burlaga_intermittent_1991,tu_mhd_1995,frisch_turbulence_1995}, the analysis of our statistically robust, large dataset quantifies these characteristics in a novel way, allowing us to treat separately stable and unstable intervals in the data. Calculation from observations of all the quantities in this study requires the use of incremental gradients, and we benefit from Solar Orbiter's in-situ instrument suite that provides continuous high-resolution datasets over significant timescales to enable this. The scientific impacts of the efficiency and high resolution of these instruments are by now well documented \citep{rouillard_models_2020,damicis_first_2021,louarn_multiscale_2021,opie_conditions_2022,zouganelis2022solar}. Nevertheless, our measures are restricted by the use of single-spacecraft data, which confines all analyses in this work to the radial sampling direction. In our view, the most promising route to extend this work lies in the use of high-quality three-dimensional data from a multi-spacecraft mission. This extension would allow the estimation of all relevant three-dimensional gradients rather than sampling along the radial direction via Taylor's hypothesis only. It would also be useful to extend the range of radial distances from the Sun over which these analyses are made using the perihelia data from Solar Orbiter and Parker Solar Probe. 

Previous work shows how unstable intervals are statistically spatially and temporally distributed \citep{opie_conditions_2022} and that unstable intervals are ergodicity-breaking and therefore statistically disjoint with respect to the stable regime \citep{opie_effect_2023}. Our current findings support and extend these previous results and describe more fully the impact of turbulence on kinetic instabilities in the solar wind, and in particular quantify the role of $\Gamma_R$ in driving unstable conditions.

\section{Conclusions}
\label{conc}

In this work, we show that solar-wind intervals with parameters above the thresholds for temperature-anisotropy-driven instabilities are on average characterised by high absolute values of $\Gamma_R$ which is a measure of the extent to which bulk motions in the plasma drive temperature anisotropy. We attribute this observational result to the proposition that strong velocity shears drive temperature anisotropies in the turbulent solar wind through the shearing of the frozen-in magnetic field with a double-adiabatic impact on the particle distributions.

The radial rate of strain $\Gamma_R$ is highly intermittent with a distribution that exhibits greater levels of kurtosis relative to those of $\boldsymbol{B}$ and $\boldsymbol{V}$ which is not unexpected given the contribution of increments of both $\boldsymbol{B}$ and $\boldsymbol{V}$ in Equation~\ref{RRS}. Nonetheless, we attribute this observational result to the burstiness of velocity shears in the solar wind with a significant occurrence rate of extreme values leading to a heavy-tailed distribution of $\Gamma_R$. The conditioned subsets of $\Gamma_R$ that relate to stable, oblique firehose unstable, and mirror-mode unstable intervals comprise distributions that are characteristically asymmetric (or symmetric in the case of stable data). The alternation of these distinct subsets contributes to the intermittency of $\Gamma_R$. We attribute this observational result to the inhomogeneity that distinguishes each of the three regions of $T_{\perp}/T_{\parallel}$--$\beta_{\parallel}$ parameter space represented by the subsets and that are both statistically and physically disjoint. Our observational measures -- skewness and kurtosis -- exhibit power-law scalings with an exponent that is consistent with the known intermittency in solar-wind turbulence.      

Our study opens several areas of extended interest that deserve further exploration. For turbulence studies, our results emphasise the importance of the velocity field for the temperature anisotropy (and potentially other kinetic properties) of the plasma. We suggest that  an analysis of the Local Energy Transfer (LET) measure derived from the Politano--Pouquet scaling law \citep{politano_von_1998,sorriso-valvo_local_2018} revised to account for pressure-anisotropic magnetohydrodynamic turbulence \citep{simon_exact_2022} would be a useful extension of the current study. In addition, the role of $\Gamma_R$ in solar-wind turbulence could be further investigated in the simulation domain, ideally building on the existing work that shows how fluctuations in the velocity field are related to anisotropy \citep{franci_high-resolution_2015,franci_two-dimensional_2016,franci_solar_2018, hellinger_turbulence_2019}. 

Our work highlights new science opportunities for multi-spacecraft missions such as HelioSwarm \citep{klein_helioswarm_2023} and Plasma Observatory \citep{retino_particle_2022}, which will provide a means of measuring fully three-dimensional rates of strain and intermittency. These data will enable a more precise unpicking of the contributions of non-compressive and compressive fluctuations as well as intermittency to the creation and modulation of temperature anisotropy. A mission like HelioSwarm or Plasma Observatory with their complements of fields and plasma instrumentation has thus the potential to create breakthroughs in our understanding of the interplay between turbulence and kinetic instabilities in space plasmas. In particular, our results suggest that accurate and high-resolution measurements of the velocity field will be of fundamental importance to this task.  

Our investigation of Solar Orbiter observations reveals a consistent picture of where and under what conditions kinetic instabilities act at the relevant scales in the turbulent solar wind. Unstable intervals are located in regions of strong velocity shear embedded in rapidly changing structures in the intermittent turbulence of the velocity field. The velocity shear and its associated impact on the frozen-in magnetic field drives and modulates the temperature anisotropy necessary to create the unstable conditions. The plasma can only remain above the instability thresholds for as long as the shears are sufficient to overcome the relaxation through the instabilities. This process takes place predominantly at the shear layers of highly intermittent structures whose distribution is strongly non-Gaussian. Accordingly, the presence of instabilities is not evenly distributed either spatially or temporally in the solar wind plasma.

\appendix
\section{Alfv\'enicity in the solar wind}

\label{appAlf}

For this analysis of the Alfv\'enicity, we consider the fluctuations on the scale of 2\,min, which is at the small-scale end of the inertial range and captures the relevant scales for the persistence of unstable intervals \citep{opie_conditions_2022}. We  define
\begin{equation}
\label{app1}
    \delta\boldsymbol{b} = \boldsymbol{B} - \|\boldsymbol{B_0}\|
\end{equation}
and
\begin{equation}
\label{app2}
    \delta\boldsymbol{v} = \boldsymbol{V} - \|\boldsymbol{V_0}\|,
\end{equation}
where $\|\cdot\|$ denotes the time average taken over a 2-minute interval centred on the time of the measurement of $\delta\boldsymbol{b}$ and $\delta\boldsymbol{v}$.

We define  the normalised cross-helicity as
\begin{equation}
\label{app3}
\sigma_c=\frac{2\left(\delta\boldsymbol{v}\bcdot\delta\boldsymbol{b}\right)}{\left(|\delta\boldsymbol{v}|^2+ |\delta\boldsymbol{b}|^2\right)}
\end{equation}
and the Alfv\'en ratio as
\begin{equation}
\label{app4}
    R_A=\frac{|\delta{\boldsymbol{v}}|^2}{|\delta{\boldsymbol{b}}|^2}.
\end{equation}
With these definitions, we create point-wise datasets for $\sigma_c$ and $R_A$ that allow us to investigate the correlations of the magnetic field and velocity field fluctuations at the scale of 2\,min.

\begin{figure}
  \centering
  \includegraphics{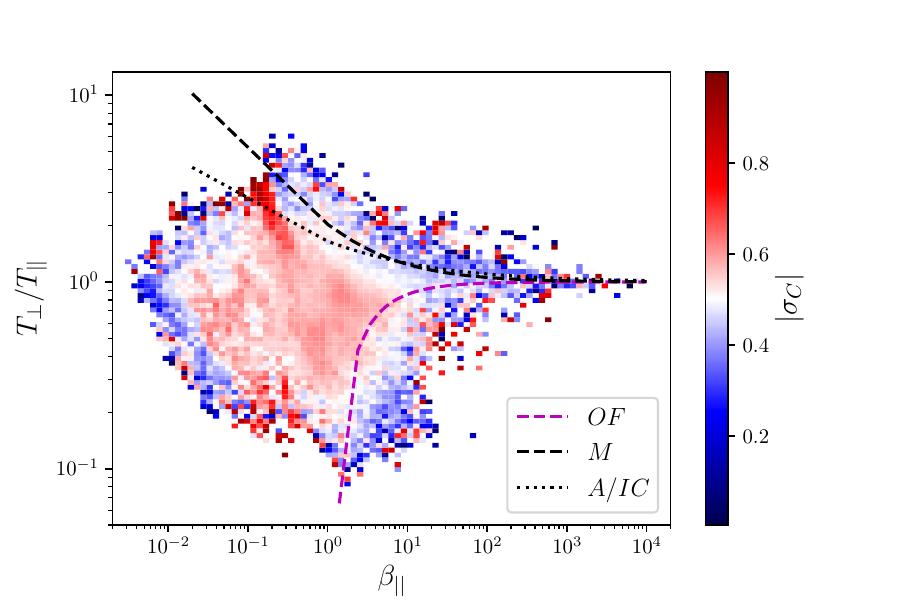}
  \caption{Distribution of $|\sigma_c|$ as bin averages in $T_{\perp}/T_{\parallel}$--$\beta_{\parallel}$ parameter space. We overplot the instability thresholds for Oblique Firehose (OF), Alfv\'en/Ion cyclotron (A/IC), and Mirror-mode (M) instabilities.}
\label{fig:app1}
\end{figure}

\begin{figure}
  \centering
  \includegraphics{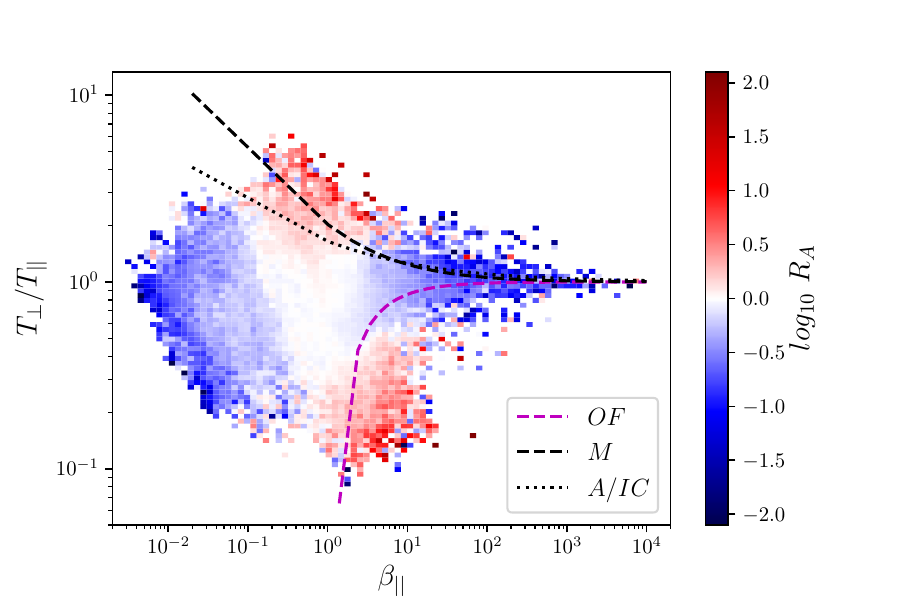}
  \caption{Distribution of $R_A$ plotted as bin averages in $T_{\perp}/T_{\parallel}$--$\beta_{\parallel}$ parameter space with instability thresholds shown for Oblique Firehose (OF), Alfv\'en/Ion cyclotron (A/IC) and Mirror-mode (M) instabilities.}
\label{fig:app2}
\end{figure}

We show $|\sigma_c|$ and $R_A$ as bin-averaged distributions in $T_{\perp}/T_{\parallel}$--$\beta_{\parallel}$ parameter space for our whole dataset in Figures~\ref{fig:app1} and \ref{fig:app2}. 

On average, normalised cross-helicity in the solar wind is higher near the core of the stable dataset in this parameter space and lower in the unstable intervals and at the low-$\beta_{\parallel}$  boundary of the stable data. The distribution of the bin-averaged Alfv\'en ratio shows that $|\delta{\boldsymbol{v}}|^2\gg|\delta{\boldsymbol{b}}|^2$ in the unstable regions of parameter space when $T_{\perp}/T_{\parallel}\leq0.6$ or $T_{\perp}/T_{\parallel}\geq1.6$.

We see that $\sigma_c$ is significantly lower on average in the unstable intervals relative to the stable data, suggesting an imbalance between the velocity and magnetic field fluctuations in the unstable regions of the $T_{\perp}/T_{\parallel}$--$\beta_{\parallel}$ parameter space. This is consistent with the hypothesis of a restriction on the amplitude of Alfv\'enic fluctuations in the limit of the oblique firehose instability \citep{squire_stringent_2016,squire_amplitude_2017} and suggests a similar mechanism may apply in the limit of the mirror-mode instability. The distribution of bin-averaged $R_A$ demonstrates that in the unstable intervals greater temperature anisotropy is associated with the dominance of energy in the velocity field over energy in the magnetic field so that $R_A\gg1$.

\section*{Declaration of Interests}

Declaration of Interests. The authors report no conflict of interest.

\section*{Acknowledgements}

S.O. is supported by NERC grant NE/S007229/1.
D.V.~and C.J.O. are supported by STFC Consolidated Grant ST/W001004/1. CHKC is supported by UKRI Future Leaders Fellowship MR/W007657/1 and STFC Consolidated Grants ST/T00018X/1 and ST/X000974/1. P.A.I. is supported by NSF grant AGS2005982. L.F. is supported by the Royal Society University Research Fellowship No. URF/R1/231710 and by UKRI/STFC grant ST/W001071/1. This research was discussed at the International Space Science Institute (ISSI) in Bern, through ISSI International Team project \#563 (Ion Kinetic Instabilities in the Solar Wind in Light of Parker Solar Probe and Solar Orbiter Observations) led by L.~Ofman and L.~Jian. Solar Orbiter is a space mission of international collaboration between ESA and NASA, operated by ESA. Solar Orbiter Solar Wind Analyser (SWA) data are derived from scientific sensors which have been designed and created, and are operated under funding provided in numerous con- tracts from the UK Space Agency (UKSA), STFC, the Agenzia Spaziale Italiana (ASI), the Centre National d'\'Etudes Spatiales (CNES), the Centre National de la Recherche Scientifique (CNRS), the Czech contribution to the ESA PRODEX programme, and NASA. Solar Orbiter SWA work at UCL/MSSL is currently funded under STFC grants ST/T001356/1 and ST/S000240/1. The Solar Orbiter magnetometer was funded by UKSA grant ST/T001062/1. 

\bibliographystyle{jpp}

\bibliography{paper3theaction}

\end{document}